\documentclass[prl,twocolumn,showpacs,groupedaddress]{revtex4}
\usepackage[dvips]{epsfig}
\usepackage{subfigure}
\usepackage{float}
\usepackage{amsmath}
\usepackage{amssymb}
\usepackage{amsfonts}
\usepackage{rotating}
\usepackage{euscript}
\usepackage{subfigure}
\usepackage{enumerate}
\usepackage{hhline}
\usepackage{threeparttable}
\usepackage{supertabular}
\usepackage{pslatex}
\usepackage{hhline}
\usepackage{threeparttable}
\usepackage{supertabular}
\usepackage{multirow}
\usepackage{array}

\usepackage{tabularx}
\usepackage{rotating}

\newcolumntype{Y}{>{\centering\arraybackslash}X}

\begin{document}

\bibliographystyle{apsrev}

\title{\bf{Fabrication-tolerant high quality factor photonic crystal microcavities}}

\author{Kartik Srinivasan}
\author{Paul E. Barclay}
\author{Oskar Painter}
\affiliation{Department of Applied Physics, California Institute of Technology, Pasadena, CA 91125, USA.}
\email{phone: (626) 395-6269, fax: (626) 795-7258, e-mail: kartik@caltech.edu}
\date{\today}

\begin{abstract}  
A two-dimensional photonic crystal microcavity design supporting a wavelength-scale volume resonant mode with a calculated quality factor ($Q$) insensitive to deviations in the cavity geometry at the level of $Q \gtrsim 2 \times 10^4$ is presented.  The robustness of the cavity design is confirmed by optical fiber-based measurements of passive cavities fabricated in silicon.  For microcavities operating in the $\lambda=1500$ nm wavelength band, quality factors between $1.3$-$4.0 \times 10^4$ are measured for significant variations in cavity geometry and for resonant mode normalized frequencies shifted by as much as $10 \%$ of the nominal value.     
\end{abstract}

\maketitle

Two-dimensional photonic crystal (PC) slab waveguide microcavities\cite{ref:Painter3,ref:Yoshie2} offer the promise of simultaneously exhibiting a high quality factor ($Q$) and an ultra-small, wavelength-scale modal volume ($V_{\text{eff}}$).  These two parameters, which physically represent long photon lifetimes and large per photon electric field strengths, respectively, are key to microcavity-enhanced processes in nonlinear optics, quantum optics, and laser physics\cite{ref:Chang,ref:Kimble2,ref:Michler,ref:Santori}.  Recent progress on PC microcavities has included theoretical work on the design of PC microcavities with predicted $Q$ factors from $10^4$ to $10^6$\cite{ref:Vuckovic2,ref:Srinivasan1,ref:Ryu5}, and experimental work demonstrating $Q$ factors in excess of $10^4$ in InP-based lasers\cite{ref:Srinivasan4} and silicon membranes\cite{ref:Akahane2,ref:Srinivasan7}.  A range of microcavity designs have been employed in these studies, and in many cases, the experimental achievement of high-$Q$ is predicated on the ability to fabricate the design with a small margin for error.  For example, in Ref. \cite{ref:Yoshie2}, the discrepancy between the fabricated device and the intended design led to a theoretical degradation of $Q$ from $3.0 \times 10^4$ to $4.4 \times 10^3$, consistent with the measured $Q$ of $2.8 \times 10^3$.  Extraordinary control over fabricated geometries has been demonstrated in recent work\cite{ref:Akahane2}, where a shift of $\sim 60$ nm in the positions of holes surrounding the cavity defect region reduced $Q$s as high as $4.5 \times 10^4$ by over an order of magnitude.  Here, we discuss work on a PC microcavity\cite{ref:Srinivasan4,ref:Srinivasan7} that exhibits a degree of robustness, both theoretically and experimentally, to deviations from the nominal design sufficient for $Q$s above $10^4$ to be maintained.  This robustness in $Q$ to changes in the PC cavity geometry is of practical importance for future experiments in the aforementioned disciplines, to provide insensitivity to fabrication imperfections, as well as to maintain the flexibility in cavity design required to form resonant modes with a prescribed field pattern and polarization.      

\begin{figure}
\begin{center}
\epsfig{figure=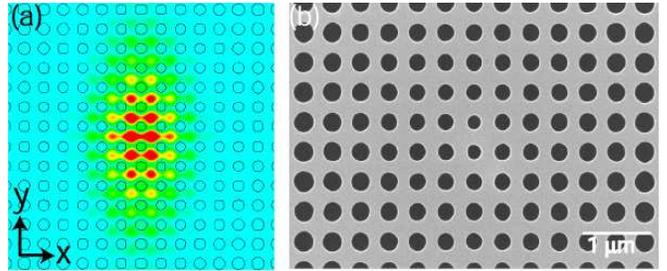, width=\linewidth}
\caption{(a) FDTD calculated magnetic field amplitude ($|\bf{B}|$) in the center of the optically thin membrane for the fundamental $A^{0}_{2}$ mode. (b) Scanning electron microscope image of a fabricated Si PC microcavity with a graded defect design (PC-5 described below).}
\label{fig:nom_PC_design}
\end{center}
\end{figure}

Radiative losses in planar waveguide two-dimensional PC defect microcavities can be separated into in-plane and out-of-plane components, quantified by the quality factors $Q_{\parallel}$ and $Q_{\perp}$, respectively, with the total radiative $Q$ given by $Q^{-1}=Q_{\parallel}^{-1}+Q_{\perp}^{-1}$.  $Q_{\parallel}$ is determined by the size and angular extent (in-plane) of the photonic bandgap, while $Q_{\perp}$ is determined by the presence of in-plane momentum components ($\bf{k}$) within the waveguide cladding light cone, which are not confined by total internal reflection at the core-cladding interface.  In Ref. \cite{ref:Srinivasan1}, PC microcavities were designed using two mechanisms to avoid radiative loss: (i) use of a mode that is odd about mirror planes normal to its dominant Fourier components, in order to eliminate the DC ($\bf{k}=0$) part of the in-plane spatial frequency spectrum and hence reduce vertical radiation loss, and (ii) use of a grade in the hole radius to further confine the mode and reduce in-plane radiative losses.  The resulting PC microcavity design within the square lattice creates a TE-like (magnetic field predominatly along $\hat{z}$) donor-type defect mode (labeled $A^{0}_{2}$\footnote{This label refers to the mode's symmetry classification and to it being the lowest frequency mode in the bandgap.}) as shown in Fig. \ref{fig:nom_PC_design}(a).  FDTD simulations of this resonant mode predict a $Q$-factor of $10^5$ and an effective modal volume of $V_{\text{eff}} \sim 1.2(\lambda/n)^3$.  We now show how use of mechanisms (i) and (ii) above create a level of robustness in the cavity design.    

Use of an odd symmetry mode to suppress vertical radiation loss is, at a basic level, independent of changes in the size of the holes defining the defect cavity.  This feature has been confirmed in simulations of simple defect cavity designs in square lattice photonic crystals\cite{ref:Srinivasan1}, where $Q_{\perp}$ did not degrade below $10^4$, despite significant changes (as much as 40$\%$) in the size of the (two) central defect holes. Perturbations that cause the cavity to be asymmetric create a mode which, though not strictly odd, will be a perturbation to an odd mode, and hence will still largely suppress DC Fourier components and exhibit high $Q$.  However, for the square lattice photonic crystal structures considered here, perturbations to the central defect hole geometry can result in a degradation in $Q_{\parallel}$, due in part to the lack of a complete in-plane bandgap within the square lattice.  This lack of a complete bandgap requires the defect geometry to be tailored so as to eliminate the presence of Fourier components in directions where the lattice is no longer reflective.  

\begin{figure}
\begin{center}
\epsfig{figure=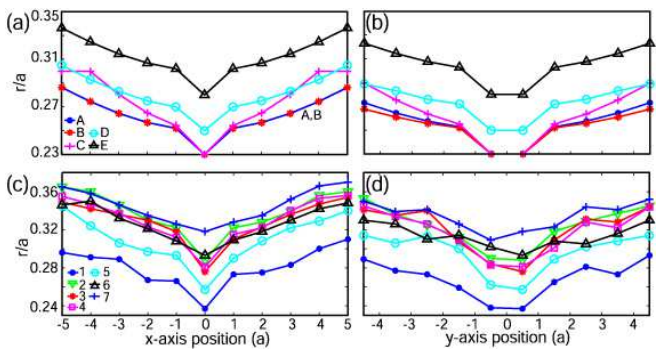, width=\linewidth}
\caption{Grade in the normalized hole radius ($r/a$) along the central $\hat{x}$ and $\hat{y}$ axes of square lattice PC cavities such as those shown in Fig. \ref{fig:nom_PC_design}. Cavity $r/a$ profiles for (a,b) FDTD cavity designs and (c,d) microfabricated Si cavities.}
\label{fig:r_a_profiles}
\end{center}
\end{figure}

\renewcommand{\arraystretch}{1.0} 
\renewcommand{\extrarowheight}{3pt}
\begin{table}
\caption{Theoretical (PC-A through PC-E) and experimental (PC-1 through PC-7) normalized frequency ($a/\lambda_o$) and quality factor ($Q$) values for the $A^{0}_{2}$ mode of cavities with profiles shown in Figure \ref{fig:r_a_profiles}.}
\label{table:A_2_mode_props}
\begin{center}
%\begin{tabular*}{\linewidth}{c|c|c|c|c|c}
\begin{tabularx}{\linewidth}{YYYYYY}
\hline
\hline
%\multicolumn{2}{|c|}{\epsfig{figure=r_a_x_FDTD_rev4.eps, width=0.45\linewidth}} & \multicolumn{3}{c|}{\epsfig{figure=r_a_y_FDTD_rev4.eps, width=0.45\linewidth}} \\
%\hhline{|=:=:=:=:=:|}
Cavity & $d/a$ & $a/\lambda_{0}$ & $Q_{\perp}$  & $Q_{\parallel}$  & $Q$ \\
\hline
PC-A & 0.750 & 0.245 & $1.1 \times 10^5$ & $4.7 \times 10^5$ & $9.0 \times 10^4$ \\
PC-B & 0.750 & 0.245 & $1.1 \times 10^5$ & $2.6 \times 10^5$ & $7.5 \times 10^4$ \\
PC-C & 0.750 & 0.247 & $1.0 \times 10^5$ & $3.7 \times 10^5$ & $8.0 \times 10^4$ \\
PC-D & 0.750 & 0.253 & $8.6 \times 10^4$ & $3.0 \times 10^5$ & $6.7 \times 10^4$ \\ 
PC-E & 0.750 & 0.266 & $6.2 \times 10^4$ & $6.5 \times 10^5$ & $5.6 \times 10^4$ \\
\hline
PC-1 & 0.879 & 0.241 & - & - & $1.6 \times 10^4$ \\
PC-2 & 0.850 & 0.255 & - & - & $1.8 \times 10^4$ \\
PC-3 & 0.850 & 0.251 & - & - & $1.7 \times 10^4$ \\
PC-4 & 0.842 & 0.251 & - & - & $2.4 \times 10^4$ \\
PC-5 & 0.842 & 0.249 & - & - & $2.5 \times 10^4$ \\
PC-6 & 0.800 & 0.263 & - & - & $4.0 \times 10^4$ \\
PC-7 & 0.800 & 0.270 & - & - & $1.3 \times 10^4$ \\
\hline
\hline
\end{tabularx}
\end{center}
\end{table}
\renewcommand{\arraystretch}{1.0} 

This tailoring was achieved in Ref. \cite{ref:Srinivasan1} by a grade in the hole radius moving from the center of the cavity outwards.  The grade, shown in Fig. \ref{fig:nom_PC_design}, serves to help eliminate couplings to in-plane radiation modes along the diagonal axes of the square lattice (the $M$-point of the reciprocal lattice) where the PC is no longer highly reflective, while simultaneously providing a means to keep the in-plane reflectivity high along the $\hat{y}$ axis (the direction of the mode's dominant Fourier components).  The use of a large number of holes to define the defect region ensures that no single hole is responsible for creating the potential well that confines the resonant mode, making the design less susceptible to fluctuations in the size of individual holes.  Instead, the continuous change in the position of the conduction band-edge resulting from the grade in hole radius creates an approximately harmonic potential well\cite{ref:Painter14}.  This smooth change in the position of the band-edge creates a robust way to mode-match between the central portion of the cavity (where the mode sits) and its exterior.  In other work\cite{ref:Akahane2}, softening of this transition is achieved by adjusting the position of two holes surrounding the central cavity region (which consists of three removed air holes in a hexagonal lattice).  This method can achieve high-$Q$, but as mode-matching is achieved by tailoring only two holes it is more sensitive to perturbations than the adiabatic transition created by a grade in the hole radius.  Finally, we note that even though a relatively large number of holes are modified to create the graded lattice, $V_{\text{eff}}$ is $\emph{still}$ wavelength-scale, and remains between $0.8$-$1.4(\lambda/n)^3$ in all of the devices considered in this work.  In addition, the methods used here to achieve robustness in $Q$ are general and can be applied to cavities in other PC lattices\cite{ref:Srinivasan2}. 

To highlight these ideas, 3D FDTD simulations of cavities  with varying grades and average normalized hole radius ($\bar{r}/a$) were performed.  Figure \ref{fig:r_a_profiles}(a)-(b) shows the grade in $r/a$ along the central $\hat{x}$ and $\hat{y}$ axes for several designs (PC-A through PC-E), and Table \ref{table:A_2_mode_props} lists the calculated resonant frequency, vertical, in-plane, and total $Q$ factors.  In all of these simulations, $Q_{\perp}$ remains close to $10^5$, with PC-E showing more significant degradation largely as a result of the increased modal frequency (creating a larger-sized cladding light cone).  In addition, an inappropriate choice of grade along the $\hat{x}$-axis can lead to increased in-plane losses via coupling to $M$-point modes.  Nevertheless, the loss in any of the simulated devices did not cause $Q$ to be reduced below $2 \times 10^4$.

\begin{figure}
\begin{center}
\epsfig{figure=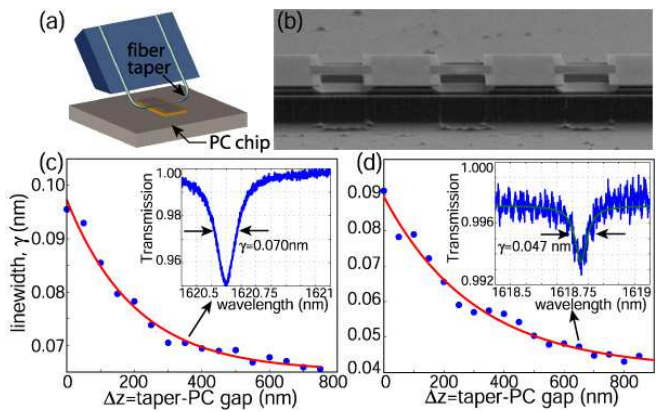, width=\linewidth}
\caption{(a) Schematic illustrating the fiber taper probe measurement setup. (b) SEM image of an array of undercut PC cavities. (c) Measured data (blue dots) and exponential fit (red curve) for linewidth vs. taper-PC gap of the $A_{2}^{0}$ mode in PC-5.  (Inset) Taper transmission for this device when the taper-PC gap is 350 nm. (d) Same as (c) for PC-6 (here, the taper transmission in the inset is shown when ${\Delta}z$=650 nm).  The transmission curves are normalized relative to transmission in the absence of the PC cavity.}
\label{fig:PC_cavity_linewidth_measurements}
\end{center}
\end{figure}

To test the sensitivity of the design to perturbations experimentally, cavities were fabricated in a $d$=340 nm thick silicon membrane through a combination of electron beam lithography, inductively-coupled plasma reactive ion etching, and wet etching.  Figure \ref{fig:r_a_profiles}(c)-(d) shows the values of $r/a$ along the central $\hat{x}$ and $\hat{y}$ axes for a number of fabricated devices (PC-1 through PC-7), as measured with a scanning electron microscope (SEM).  Cavities are passively tested\cite{ref:Srinivasan7} using an optical fiber taper\cite{ref:Knight}, which consists of a standard single mode optical fiber that has been heated and stretched to a minimum diameter of $\text{1-2}$ $\mu$m.  At such sizes, the evanescent field of the fiber mode extends into the surrounding air, providing a means by which the cavity modes can be sourced and out-coupled. The fiber taper is spliced to the output of a fiber-pigtailed scanning tunable laser (1565-1625 nm) with 1 pm resolution, and is mounted (Fig. \ref{fig:PC_cavity_linewidth_measurements}(a)) above and parallel to an array of PC cavities (Fig. \ref{fig:PC_cavity_linewidth_measurements}(b)).  When it is brought into close proximity ($\sim 500$ nm) to the sample surface, evanescent coupling between the taper and cavity modes occurs.             

Fig. \ref{fig:PC_cavity_linewidth_measurements}(c)-(d) shows measurements for devices PC-5 and PC-6, which have significantly different $r/a$ profiles (Figure \ref{fig:r_a_profiles}(c)-(d)).  The inset of Fig. \ref{fig:PC_cavity_linewidth_measurements}(c) shows the normalized taper transmission as a function of wavelength when the taper is $350$ nm above cavity PC-5.  By measuring the dependence of cavity mode linewidth ($\gamma$) on the vertical taper-PC gap (${\Delta}z$) (Fig. \ref{fig:PC_cavity_linewidth_measurements}(c)), an estimate of the true cold-cavity linewidth ($\gamma_{0}$) is given by the asymptotic value of $\gamma$ reached when the taper is far from the cavity.  For PC-5, $\gamma_{0} \sim 0.065$ nm, corresponding to $Q \sim 2.5 \times 10^4$.  Fig. \ref{fig:PC_cavity_linewidth_measurements}(d) shows the linewidth measurement for PC-6.  For this device, $\gamma_{0} \sim 0.041$ nm, corresponding to a $Q \sim 4.0 \times 10^4$.  As described in Ref. \cite{ref:Srinivasan7}, the strength of the taper-PC coupling as a function of taper position can be used to estimate the spatial localization of the cavity field; these measurements closely correspond with calculations and for PC-6 are consistent with an FDTD-predicted $V_{\text{eff}} \sim 0.9(\lambda/n)^3$.  These PC microcavities thus simultaneously exhibit a high-$Q$ factor that is insensitive to perturbations, and an ultra-small $V_{\text{eff}}$.    

Linewidth measurements for each of the cavities PC-1 through PC-7 are compiled in Table \ref{table:A_2_mode_props}.  The robustness of the $Q$ to non-idealities in fabrication is clearly evident.  Though all of the devices exhibit a general grade in $r/a$, the steepness of the grade and the average hole radius ($\bar{r}/a$) vary considerably without reducing $Q$ below $1.3 \times 10^4$.  These high-$Q$ values are exhibited despite the fact that many cavities are not symmetric (the odd boundary condition is thus only approximately maintained), and the frequency of the cavity resonance varies over a $10\%$ range, between $a/\lambda_{o} = 0.243$-$0.270$.  

The measured $Q$ values in Table \ref{table:A_2_mode_props} are still lower than predicted from simulations.  This discrepancy is likely due in part to slightly angled etched sidewalls that have been shown in calculations to lead to radiative coupling to TM-like modes\cite{ref:Tanaka}.  This non-ideality helps explain why PC-1, which is closest in $r/a$ value to the desired design (PC-A), does not exhibit the highest $Q$ experimentally.  In particular, we have observed that the sidewall angle is poorer for smaller sized holes.  On the other end of the spectrum, cavities with the largest hole sizes such as PC-7, which may have more vertical sidewalls, also begin to exhibit higher vertical radiation loss as a result of a larger modal frequency and cladding light cone.  In addition, surface roughness is a potential source of loss; for PC-6, which exhibited the highest $Q$ value, a chemical resist stripping process was used (rather than a plasma de-scum) and may have produced a cleaner, smoother surface.            

In summary, the robustness in $Q$ to errors in the in-plane design of a PC microcavity consisting of a graded square lattice of air holes is discussed.  This property is confirmed both by FDTD simulations of devices where the steepness of the grade and the average hole radius are varied without degrading $Q$ below $2 \times 10^4$, and in measurements of microfabricated Si cavities that exhibit $Q$ factors between $1.3$-$4.0 \times 10^4$ over a wide range of device parameters.  For these high-$Q$ cavities, current limitations on the $Q$ factor appear to stem principally from slightly angled sidewalls and etched surface roughness, as opposed to errors in the in-plane shape or size of holes.    

This work was partly supported by the Charles Lee Powell Foundation.  The authors thank M. Borselli for his contributions in building the taper test setup.  K.S. thanks the Hertz Foundation for its financial support.

%\bibliography{/home/kartik/PBG_bibliography/PBG}
\bibliography{./PBG}
\end{document}